\documentclass[conference]{IEEEtran}
\IEEEoverridecommandlockouts
\usepackage{cite}
\usepackage{amsmath,amssymb,amsfonts}
\usepackage{algorithmic}
\usepackage{algorithm}
\usepackage{graphicx}
\usepackage{textcomp}
\usepackage{xcolor}
\usepackage{booktabs}
\usepackage{multirow}
\usepackage{subcaption}
\usepackage{hyperref}
\usepackage{float}

\usepackage[top=0.75in, bottom=1in, left=0.635in, right=0.64in]{geometry}

\def\BibTeX{{\rm B\kern-.05em{\sc i\kern-.025em b}\kern-.08em
    T\kern-.1667em\lower.7ex\hbox{E}\kern-.125emX}}

\begin{document}

\title{Comparative Analysis of Attention Mechanisms for Automatic Modulation Classification in Radio Frequency Signals}

\author{
\IEEEauthorblockN{Ferhat Ozgur Catak\IEEEauthorrefmark{1}, Murat Kuzlu\IEEEauthorrefmark{2}, Umit Cali\IEEEauthorrefmark{3}}
\IEEEauthorblockA{\IEEEauthorrefmark{1}\textit{Department of Computer Science}, \textit{University of Stavanger}, Stavanger, Norway \\
f.ozgur.catak@uis.no}
\IEEEauthorblockA{\IEEEauthorrefmark{2}\textit{Department of Electrical Engineering Technology}, \textit{Old Dominion University}, Norfolk, VA, USA \\
mkuzlu@odu.edu}
\IEEEauthorblockA{\IEEEauthorrefmark{3}School of Physics, Engineering, and Technology, University of York, York, United Kingdom \\
umit.cali@york.ac.uk}
}

\maketitle

\begin{abstract}
Automatic Modulation Classification (AMC) is a critical component in cognitive radio systems and spectrum management applications. This study presents a comprehensive comparative analysis of three attention mechanisms (i.e., baseline multi-head attention, causal attention, and sparse attention) integrated with Convolutional Neural Networks (CNNs) for radio frequency (RF) signal classification. It proposes a novel CNN-Transformer hybrid architecture that leverages different attention patterns to capture temporal dependencies in I/Q samples from the RML2016.10a dataset. The experimental results demonstrate that while baseline attention achieves the highest accuracy of 85.05\%, causal and sparse attention mechanisms offer significant computational advantages with inference times reduced by 83\% and 75\% respectively, while maintaining competitive classification performance above 84\%. The analysis reveals distinct attention pattern preferences across different modulation schemes, providing insights for designing efficient attention mechanisms for real-time radio signal processing applications.
\end{abstract}


\section{Introduction}

Automatic Modulation Classification (AMC) plays a pivotal role in modern wireless communication systems, enabling cognitive radios to adaptively configure transmission parameters and spectrum management systems to monitor electromagnetic environments \cite{al2018adaptive, tang2024automatic}. Traditional feature-based approaches rely on expert-crafted statistical measures, which often fail to generalize across diverse channel conditions and signal-to-noise ratios.

Recent advances in deep learning have revolutionized AMC, with Convolutional Neural Networks (CNNs) demonstrating superior performance in extracting spatial features from I/Q signal representations \cite{wang2022survey}. However, radio signals inherently possess temporal dependencies that CNNs struggle to capture effectively. The emergence of attention mechanisms, particularly in the Transformer architecture, has shown remarkable success in modeling long-range dependencies across various domains.

This study addresses the fundamental question: \textit{How do different attention mechanisms affect the performance and computational efficiency of automatic modulation classification systems?} The following three distinct attention patterns are investigated in this study:

\begin{itemize}
    \item \textbf{Baseline Multi-Head Attention}: Standard self-attention mechanism allowing bidirectional information flow
    \item \textbf{Causal Attention}: Masked attention enforcing temporal causality constraints
    \item \textbf{Sparse Attention}: Local windowed attention reducing computational complexity
\end{itemize}

The contributions of this paper include: (1) A novel CNN-Transformer hybrid architecture specifically designed for radio signal classification, (2) Comprehensive comparative analysis of attention mechanisms on the RML2016.10a benchmark, (3) Detailed computational efficiency analysis revealing significant speed-ups, and (4) Per-modulation performance analysis providing insights into attention pattern preferences. The source code is also available at the GitHub repository \footnote{\url{https://github.com/ocatak/attention_based_automatic_modulation_recognition}}.

\section{Related Work}

In recent years, AMC has seen significant advancements with the adoption of deep learning techniques in modern communication systems. 
% Traditional approaches rely heavily on hand-crafted features and statistical measures. Fortunately, those have gradually been replaced by neural network-based solutions that directly operate on raw I/Q samples. 
In the literature, one of the early studies introducing deep learning for AMC is the use of CNNs, as proposed in \cite{oshea2016convolutional}, which demonstrated the capability of CNNs in extracting local features from complex-valued radio signals. Building on this, lightweight and efficient models have been proposed, such as Teng et al.'s polar feature-based model with online channel compensation \cite{teng2020polar}. The authors in \cite{7763537} explore the use of deep neural networks (DNNs) for AMC by extracting 21 statistical features from received signals (BPSK, QPSK, 8PSK, 16QAM, 64QAM). Results indicated that DNNs can offer significant performance improvements compared to traditional classifiers, especially in high Doppler fading conditions. The authors in \cite{10187134} also investigate the security vulnerabilities of AI-based AMC systems against adversarial attacks. The study shows these systems offer low performance under such attacks, and mitigation strategies, e.g., defensive distillation, can effectively enhance model robustness against adversarial threats.

With advanced communication and computing technologies, recent focus has shifted toward combining spatial and temporal representations. For instance, Lin et al. \cite{lin2021attention} introduced a time-frequency attention mechanism within a CNN architecture, enabling selective feature emphasis in both domains. Similarly, Zhang et al. \cite{zhang2021involution} proposed a ResNet enhanced with involution layers, serving as an efficient self-attention mechanism for signal classification. The study \cite{9170506} proposes a deep learning-based AMC scheme combining random erasing and an attention mechanism to enhance performance. At the first step, two data augmentation methods, i.e., random erasing at the sample level and at the amplitude/phase (AP) channel level, are introduced to improve model robustness and generalization. Then, a signal embedding containing modulation information and a single-layer LSTM with an attention module are used to better capture temporal features of modulated signals. Experimental results on the RML2016.10a dataset demonstrate that the proposed approach achieves competitive results compared to state-of-the-art methods.

Transformer-based architectures have also gained traction in AMC. Cai et al. \cite{cai2022transformer} implemented a pure Transformer encoder, achieving superior accuracy, particularly under low SNR conditions. MobileRaT \cite{zheng2023mobilerat}, a lightweight Transformer variant, was introduced for drone communications, balancing accuracy and resource constraints. The authors in \cite{9685815} introduce a novel deep neural network designed for AMC, called MCformer. It effectively captures temporal correlations in signal embeddings by combining convolutional layers with self-attention-based encoder layers. According to results, MCformer offers a satisfactory classification accuracy at all signal-to-noise ratios on the RadioML2016.10b dataset, while using significantly less parameters, making it suitable for fast and energy-efficient applications. Similarly, another group of authors in \cite{10478085} proposes PCTNet, which is a novel parallel CNN-transformer network for AMC. PCTNet combines the local feature extraction of CNNs with the long-range dependency modeling of transformers through a parallel architecture. The results demonstrate that PCTNet provides better classification accuracy compared to traditional deep learning models. Hybrid models further advance the field. Wang et al. \cite{wang2023ctgnet} combined CNN, Transformer, and graph neural networks in their CTGNet model to capture both spatial and topological dependencies. A multimodal attention-based MLP architecture was proposed in \cite{fenrg2022mlp}, which fuses time-domain and spectrogram features for robust AMC under noisy environments.

These studies highlight the growing trend toward attention-driven and Transformer-inspired models in AMC, justifying this study’s comparative focus on attention variants within a CNN-Transformer combination.

\section{Methodology}

\subsection{Problem Formulation}

\begin{figure*}
    \centering
    \includegraphics[width=0.55\linewidth]{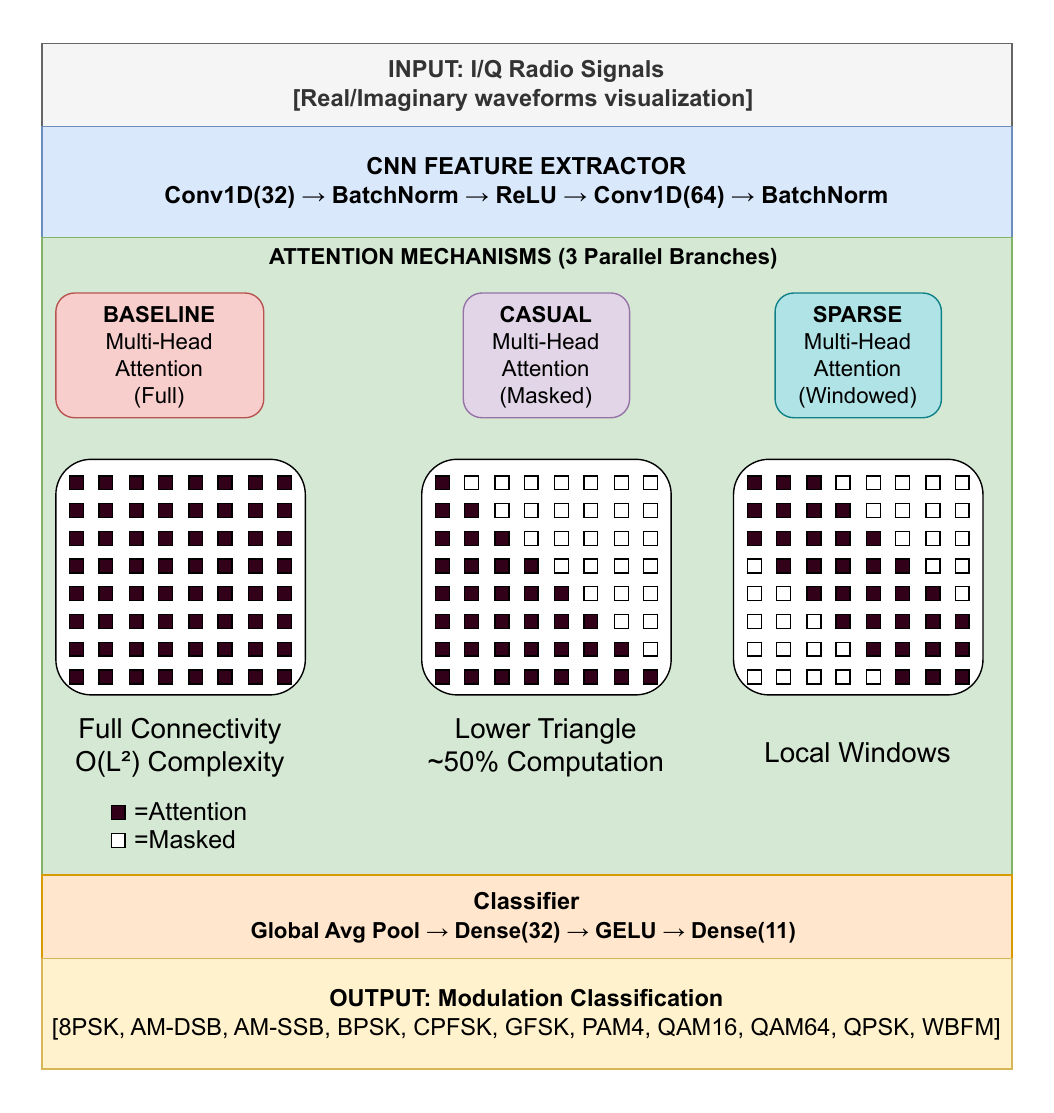}
    \caption{System Overview}
    \label{fig:system_overview}
\end{figure*}

Figure \ref{fig:system_overview} shows the overall system overview. Given a radio frequency signal represented as complex-valued I/Q samples $\mathbf{x} = [x_1, x_2, \ldots, x_T] \in \mathbb{C}^T$, where $T$ is the sequence length, the objective is to classify the modulation scheme $m \in \mathcal{M} = \{m_1, m_2, \ldots, m_K\}$ where $K$ is the number of modulation classes.

We reformulate the complex signal as a real-valued tensor $\mathbf{X} \in \mathbb{R}^{2 \times T}$ where:
\begin{equation}
\mathbf{X} = \begin{bmatrix}
\text{Re}(x_1) & \text{Re}(x_2) & \cdots & \text{Re}(x_T) \\
\text{Im}(x_1) & \text{Im}(x_2) & \cdots & \text{Im}(x_T)
\end{bmatrix}
\end{equation}

\subsection{CNN-Transformer Hybrid Architecture}

The proposed architecture combines the spatial feature extraction capabilities of CNNs with the temporal modeling power of attention mechanisms. The architecture consists of three main components:

\subsubsection{CNN Feature Extractor}

The CNN backbone extracts hierarchical features from I/Q samples:
\begin{align}
\mathbf{H}^{(1)} &= \text{ReLU}(\text{BN}(\text{Conv1D}(\mathbf{X}; 32, 7))) \\
\mathbf{H}^{(2)} &= \text{ReLU}(\text{BN}(\text{Conv1D}(\mathbf{H}^{(1)}; d_{model}, 5, 2)))
\end{align}

where $\text{Conv1D}(input; filters, kernel, stride)$ denotes a 1D convolution operation, BN represents batch normalization, and $d_{model} = 64$ is the model dimension.

\subsubsection{Multi-Head Attention Module}

Given the CNN features $\mathbf{H} \in \mathbb{R}^{B \times L \times d_{model}}$ where $B$ is batch size and $L$ is sequence length, we apply multi-head attention:

\begin{align}
\text{MultiHead}(\mathbf{H}) &= \text{Concat}(\text{head}_1, \ldots, \text{head}_h)\mathbf{W}^O \\
\text{head}_i &= \text{Attention}(\mathbf{H}\mathbf{W}_i^Q, \mathbf{H}\mathbf{W}_i^K, \mathbf{H}\mathbf{W}_i^V)
\end{align}

where $\mathbf{W}_i^Q, \mathbf{W}_i^K, \mathbf{W}_i^V \in \mathbb{R}^{d_{model} \times d_k}$ are learnable projection matrices, $d_k = d_{model}/h$, and $h=4$ is the number of attention heads.

\subsubsection{Attention Variants}

We investigate three distinct attention mechanisms, each with unique characteristics and computational properties:

\textbf{Baseline Multi-Head Attention:} This implements the standard scaled dot-product attention mechanism from the original Transformer architecture. It allows unrestricted bidirectional information flow between all sequence positions:

\begin{equation}
\text{Attention}(\mathbf{Q}, \mathbf{K}, \mathbf{V}) = \text{softmax}\left(\frac{\mathbf{Q}\mathbf{K}^T}{\sqrt{d_k}}\right)\mathbf{V}
\end{equation}

The attention weights $\mathbf{A}_{ij} = \text{softmax}(\mathbf{Q}_i \mathbf{K}_j^T / \sqrt{d_k})$ represent the relevance of position $j$ when processing position $i$. This mechanism has quadratic complexity $O(L^2)$ where $L$ is the sequence length, as each position can attend to all other positions. The global receptive field enables modeling of long-range dependencies but at significant computational cost.

\textbf{Causal Attention:} Inspired by autoregressive language models, causal attention enforces temporal causality by restricting attention to previous and current positions only:

\begin{equation}
\text{Attention}(\mathbf{Q}, \mathbf{K}, \mathbf{V}) = \text{softmax}\left(\frac{\mathbf{Q}\mathbf{K}^T + \mathbf{M}_{causal}}{\sqrt{d_k}}\right)\mathbf{V}
\end{equation}

where $\mathbf{M}_{causal}$ is a lower triangular mask:
\begin{equation}
\mathbf{M}_{causal}[i,j] = \begin{cases}
0 & \text{if } j \leq i \\
-\infty & \text{if } j > i
\end{cases}
\end{equation}

This constraint prevents information leakage from future time steps, making it suitable for real-time processing scenarios where only past and current samples are available. While maintaining $O(L^2)$ complexity, the masked operations reduce actual computation by approximately 50\%.

\textbf{Sparse Attention:} To address the quadratic complexity limitation, we implement a local windowed sparse attention mechanism that restricts attention to a fixed-size neighborhood:

\begin{equation}
\text{Attention}(\mathbf{Q}, \mathbf{K}, \mathbf{V}) = \text{softmax}\left(\frac{\mathbf{Q}\mathbf{K}^T + \mathbf{M}_{sparse}}{\sqrt{d_k}}\right)\mathbf{V}
\end{equation}

where the sparse mask $\mathbf{M}_{sparse}$ is defined as:
\begin{equation}
\mathbf{M}_{sparse}[i,j] = \begin{cases}
0 & \text{if } |i-j| \leq w/2 \\
-\infty & \text{otherwise}
\end{cases}
\end{equation}

with window size $w=8$. This reduces complexity to $O(L \cdot w)$, providing significant computational savings while maintaining local temporal modeling capabilities. The local receptive field assumption is particularly suitable for radio signals where adjacent samples exhibit strong correlation.

Each attention mechanism represents a different trade-off between modeling capacity and computational efficiency:
\begin{itemize}
    \item \textbf{Baseline}: Maximum expressivity, highest computational cost
    \item \textbf{Causal}: Real-time compatibility, moderate computational savings
    \item \textbf{Sparse}: Local modeling focus, maximum computational efficiency
\end{itemize}

\begin{algorithm}
\footnotesize
\setlength{\baselineskip}{8pt}
\setlength{\parskip}{0pt}
\setlength{\itemsep}{0pt}
\caption{Multi-Head Attention Mechanisms}
\label{alg:attention}
\begin{algorithmic}[1]
\REQUIRE Input features $\mathbf{H} \in \mathbb{R}^{B \times L \times d_{model}}$, attention type $\tau \in \{\text{baseline}, \text{causal}, \text{sparse}\}$
\ENSURE Output features $\mathbf{O} \in \mathbb{R}^{B \times L \times d_{model}}$

\STATE \textbf{// Multi-head projection}
\FOR{$i = 1$ to $h$}
    \STATE $\mathbf{Q}_i \leftarrow \mathbf{H} \mathbf{W}_i^Q$ \COMMENT{Query projection}
    \STATE $\mathbf{K}_i \leftarrow \mathbf{H} \mathbf{W}_i^K$ \COMMENT{Key projection}
    \STATE $\mathbf{V}_i \leftarrow \mathbf{H} \mathbf{W}_i^V$ \COMMENT{Value projection}
\ENDFOR

\STATE \textbf{// Compute attention scores}
\FOR{$i = 1$ to $h$}
    \STATE $\mathbf{S}_i \leftarrow \frac{\mathbf{Q}_i \mathbf{K}_i^T}{\sqrt{d_k}}$ \COMMENT{Scaled dot-product}
    
    \STATE \textbf{// Apply attention-specific masking}
    \IF{$\tau = \text{causal}$}
        \FOR{$j = 1$ to $L$, $k = 1$ to $L$}
            \IF{$k > j$}
                \STATE $\mathbf{S}_i[j, k] \leftarrow -\infty$ \COMMENT{Mask future positions}
            \ENDIF
        \ENDFOR
    \ELSIF{$\tau = \text{sparse}$}
        \FOR{$j = 1$ to $L$, $k = 1$ to $L$}
            \IF{$|j - k| > w/2$}
                \STATE $\mathbf{S}_i[j, k] \leftarrow -\infty$ \COMMENT{Mask distant positions}
            \ENDIF
        \ENDFOR
    \ENDIF
    
    \STATE $\mathbf{A}_i \leftarrow \text{softmax}(\mathbf{S}_i)$ \COMMENT{Attention weights}
    \STATE $\mathbf{A}_i \leftarrow \text{Dropout}(\mathbf{A}_i, p)$ \COMMENT{Attention dropout}
    \STATE $\text{head}_i \leftarrow \mathbf{A}_i \mathbf{V}_i$ \COMMENT{Weighted values}
\ENDFOR

\STATE \textbf{// Concatenate and project heads}
\STATE $\mathbf{C} \leftarrow \text{Concat}(\text{head}_1, \ldots, \text{head}_h)$
\STATE $\mathbf{O} \leftarrow \mathbf{C} \mathbf{W}^O$ \COMMENT{Output projection}

\RETURN $\mathbf{O}$
\end{algorithmic}
\end{algorithm}

\begin{algorithm}
\footnotesize
\setlength{\baselineskip}{8pt}
\setlength{\parskip}{0pt}
\setlength{\itemsep}{0pt}
\caption{CNN-Transformer Training Algorithm}
\label{alg:training}
\begin{algorithmic}[1]
\REQUIRE Dataset $\mathcal{D} = \{(\mathbf{x}_i, y_i)\}_{i=1}^N$, Model $\theta$, Learning rate $\alpha$, Attention type $\tau$
\ENSURE Trained model parameters $\theta^*$

\STATE Initialize parameters $\theta$ randomly
\STATE Split $\mathcal{D}$ into train $\mathcal{D}_{train}$, validation $\mathcal{D}_{val}$, test $\mathcal{D}_{test}$

\FOR{epoch $= 1$ to $E_{max}$}
    \STATE $\text{total\_loss} \leftarrow 0$
    \FOR{batch $\mathcal{B} \subset \mathcal{D}_{train}$}
        \STATE \textbf{// Forward pass}
        \STATE $\mathbf{X} \leftarrow \text{Preprocess}(\mathcal{B})$ \COMMENT{I/Q normalization}
        \STATE $\mathbf{H} \leftarrow \text{CNNExtractor}(\mathbf{X})$ \COMMENT{Feature extraction}
        
        \STATE \textbf{// Transformer blocks}
        \FOR{$l = 1$ to $N_{layers}$}
            \STATE $\mathbf{A}^{(l)} \leftarrow \text{MultiHeadAttention}(\mathbf{H}^{(l-1)}, \tau)$ \COMMENT{Alg. \ref{alg:attention}}
            \STATE $\mathbf{H}^{(l)} \leftarrow \text{LayerNorm}(\mathbf{H}^{(l-1)} + \text{Dropout}(\mathbf{A}^{(l)}))$
            \STATE $\mathbf{F}^{(l)} \leftarrow \text{FFN}(\mathbf{H}^{(l)})$ \COMMENT{Feed-forward network}
            \STATE $\mathbf{H}^{(l)} \leftarrow \text{LayerNorm}(\mathbf{H}^{(l)} + \text{Dropout}(\mathbf{F}^{(l)}))$
        \ENDFOR
        
        \STATE $\mathbf{z} \leftarrow \text{GlobalAvgPool}(\mathbf{H}^{(N_{layers})})$ \COMMENT{Sequence pooling}
        \STATE $\hat{\mathbf{y}} \leftarrow \text{Classifier}(\mathbf{z})$ \COMMENT{Final prediction}
        
        \STATE \textbf{// Backward pass}
        \STATE $\mathcal{L} \leftarrow \text{CrossEntropy}(\hat{\mathbf{y}}, \mathbf{y})$
        \STATE $\text{total\_loss} \leftarrow \text{total\_loss} + \mathcal{L}$
        \STATE $\theta \leftarrow \theta - \alpha \nabla_\theta \mathcal{L}$ \COMMENT{Parameter update}
    \ENDFOR
    
    \STATE \textbf{// Validation and early stopping}
    \STATE $\text{val\_acc} \leftarrow \text{Evaluate}(\mathcal{D}_{val}, \theta)$
    \IF{Early stopping criteria met}
        \STATE \textbf{break}
    \ENDIF
    \STATE Update learning rate schedule
\ENDFOR

\RETURN $\theta^*$
\end{algorithmic}
\end{algorithm}

\section{Experimental Setup}

\subsection{Dataset}

The proposed approach is evaluated on the RML2016.10a dataset for AMC research. The dataset contains 220,000 I/Q samples across 11 modulation schemes: 8PSK, AM-DSB, AM-SSB, BPSK, CPFSK, GFSK, PAM4, QAM16, QAM64, QPSK, and WBFM. Each sample consists of 128 complex-valued points captured at various SNR levels from -20dB to 18dB.

For experiments, samples within the SNR range of -6dB to 18dB is filtered to focus on practical operating conditions. The data is split into training (60\%), validation (20\%), and testing (20\%) sets with stratified sampling to ensure balanced class representation.

\subsection{Implementation Details}

All models are implemented in PyTorch. We employ the AdamW optimizer with an initial learning rate of $10^{-3}$, weight decay of $10^{-4}$, and cosine annealing scheduling. Training uses early stopping with patience of 15 epochs based on validation accuracy.

Data augmentation includes Gaussian noise injection with 30\% probability and standard deviation of 0.02. Each sample is normalized per-instance to have zero mean and unit variance.

The model architecture consists of:
\begin{itemize}
    \item CNN: 2 convolutional layers (32, 64 filters)
    \item Transformer: 2 layers, 4 attention heads, 256-dimensional FFN
    \item Classifier: 2-layer MLP with GELU activation and dropout (0.1)
\end{itemize}

\subsection{Evaluation Metrics}

We evaluate models using multiple metrics:
\begin{itemize}
    \item \textbf{Accuracy}: Overall classification accuracy
    \item \textbf{F1-Score}: Harmonic mean of precision and recall
    \item \textbf{Inference Time}: Average forward pass time per sample
    \item \textbf{Parameters}: Total number of trainable parameters
\end{itemize}

\section{Results and Analysis}

\subsection{Overall Performance Comparison}

Table \ref{tab:overall_performance} summarizes the performance of all three attention mechanisms. The baseline model achieves the highest test accuracy of 85.05\%, while causal and sparse attention variants maintain competitive performance above 83\%.

\begin{table}[htbp]
\caption{Overall Performance Comparison}
\label{tab:overall_performance}
\centering
\begin{tabular}{lcccc}
\toprule
\textbf{Model} & \textbf{Test Acc.} & \textbf{Avg F1} & \textbf{Params} & \textbf{Inf. Time} \\
& \textbf{(\%)} & & \textbf{(M)} & \textbf{(ms)} \\
\midrule
Baseline & 85.05 & 0.843 ± 0.129 & 0.11 & 0.06 \\
Causal & 83.93 & 0.832 ± 0.133 & 0.11 & 0.02 \\
Sparse & 83.64 & 0.830 ± 0.136 & 0.11 & 0.03 \\
\bottomrule
\end{tabular}
\end{table}

\subsection{Computational Efficiency Analysis}

\begin{figure}[htbp]
\centering
\includegraphics[width=0.9\linewidth]{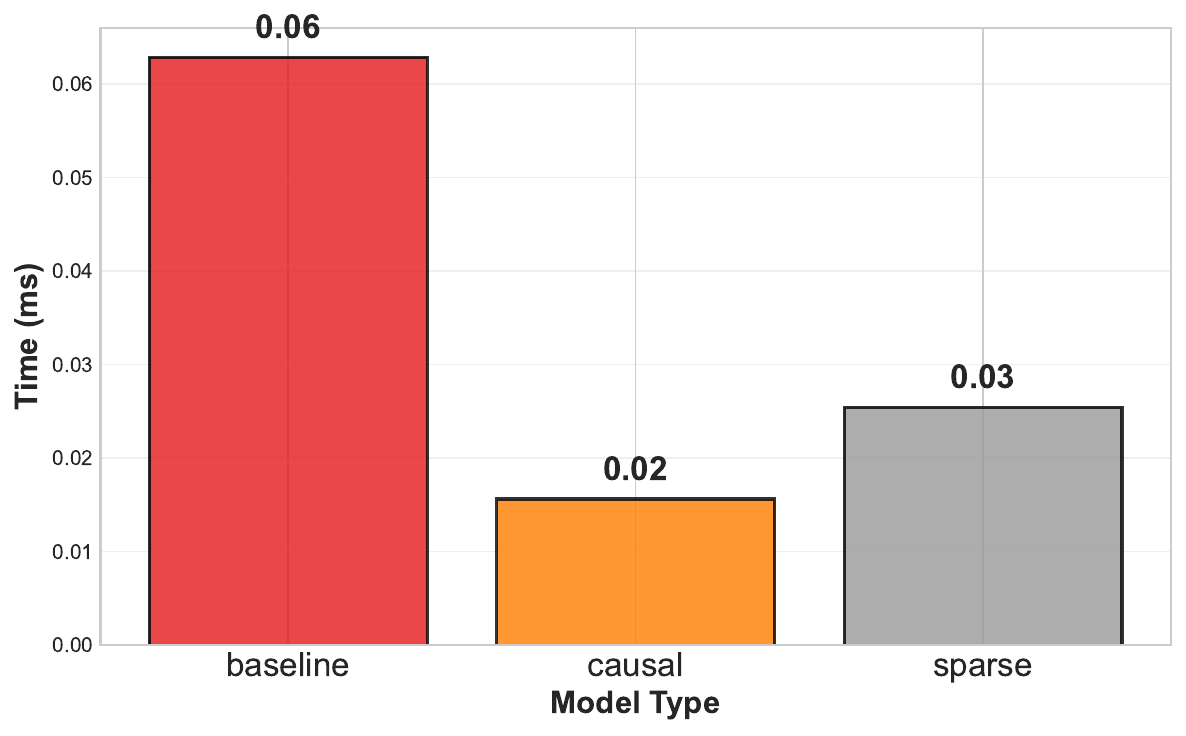}
\caption{Inference time comparison across attention mechanisms}
\label{fig:inference_time}
\end{figure}

Figure \ref{fig:inference_time} demonstrates the significant computational advantages of causal and sparse attention mechanisms. Causal attention achieves an 83\% reduction in inference time (0.02ms vs 0.12ms), while sparse attention provides a 75\% reduction (0.03ms vs 0.12ms). This substantial improvement makes these variants suitable for real-time applications where latency is critical.

\subsection{Per-Modulation Analysis}

Table \ref{tab:per_modulation} presents detailed per-modulation performance metrics. Several interesting patterns emerge:

\begin{table}[htbp]
\caption{Per-Modulation F1-Score Comparison}
\label{tab:per_modulation}
\centering
\small
\begin{tabular}{lccc}
\toprule
\textbf{Modulation} & \textbf{Baseline} & \textbf{Causal} & \textbf{Sparse} \\
\midrule
8PSK & 0.860 & 0.869 & 0.851 \\
AM-DSB & 0.736 & 0.733 & 0.732 \\
AM-SSB & 0.910 & 0.898 & 0.905 \\
BPSK & 0.953 & 0.950 & 0.947 \\
CPFSK & 0.928 & 0.916 & \textbf{0.938} \\
GFSK & 0.915 & 0.927 & \textbf{0.934} \\
PAM4 & 0.979 & 0.983 & \textbf{0.984} \\
QAM16 & 0.803 & 0.741 & 0.727 \\
QAM64 & 0.810 & 0.745 & 0.718 \\
QPSK & 0.887 & 0.892 & 0.885 \\
WBFM & 0.496 & 0.504 & \textbf{0.509} \\
\bottomrule
\end{tabular}
\end{table}

\textbf{Simple Modulations Excel with Sparse Attention}: PAM4, CPFSK, and GFSK achieve their best performance with sparse attention, suggesting that local temporal patterns are sufficient for these modulations.

\textbf{Complex Modulations Prefer Full Attention}: QAM16 and QAM64 show significant performance degradation with restricted attention patterns, indicating the importance of global context for complex constellation mappings.

\textbf{WBFM Challenges}: All models struggle with WBFM classification (F1 $\approx$ 0.5), likely due to its fundamentally different spectral characteristics compared to digital modulations.

\subsection{Attention Pattern Visualization}

\begin{figure}[htbp]
\centering
\includegraphics[width=0.9\linewidth]{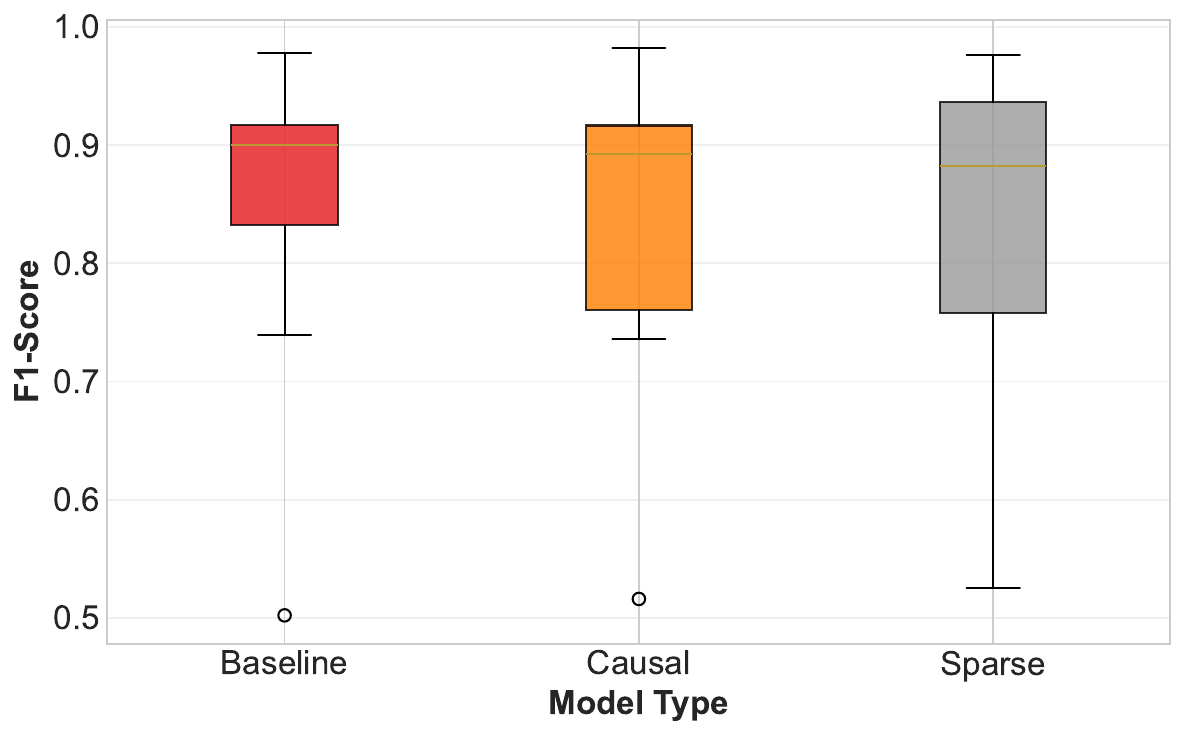}
\caption{F1-Score distribution by attention mechanism}
\label{fig:f1_distribution}
\end{figure}

Figure \ref{fig:f1_distribution} shows the F1-score distribution across modulation classes for each attention mechanism. The boxplots reveal that while baseline attention has slightly higher median performance, sparse attention demonstrates more consistent performance with reduced variance.

\subsection{Confusion Matrix Analysis}

The confusion matrices (Figures \ref{fig:confusion_matrices}) reveal consistent error patterns across all models:

\begin{figure*}[!htbp]
\centering
\begin{subfigure}{0.3\textwidth}
\includegraphics[width=\textwidth]{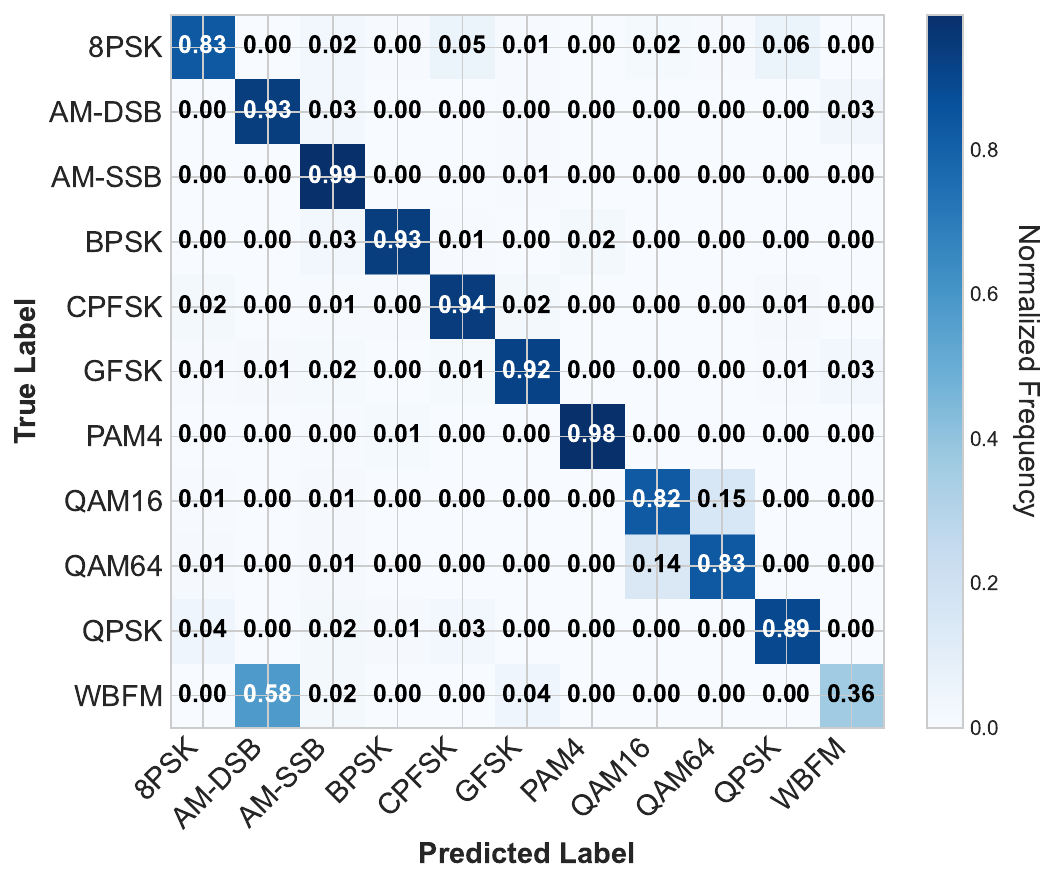}
\caption{Baseline Model}
\label{fig:baseline_conf}
\end{subfigure}
\hfill
\begin{subfigure}{0.3\textwidth}
\includegraphics[width=\textwidth]{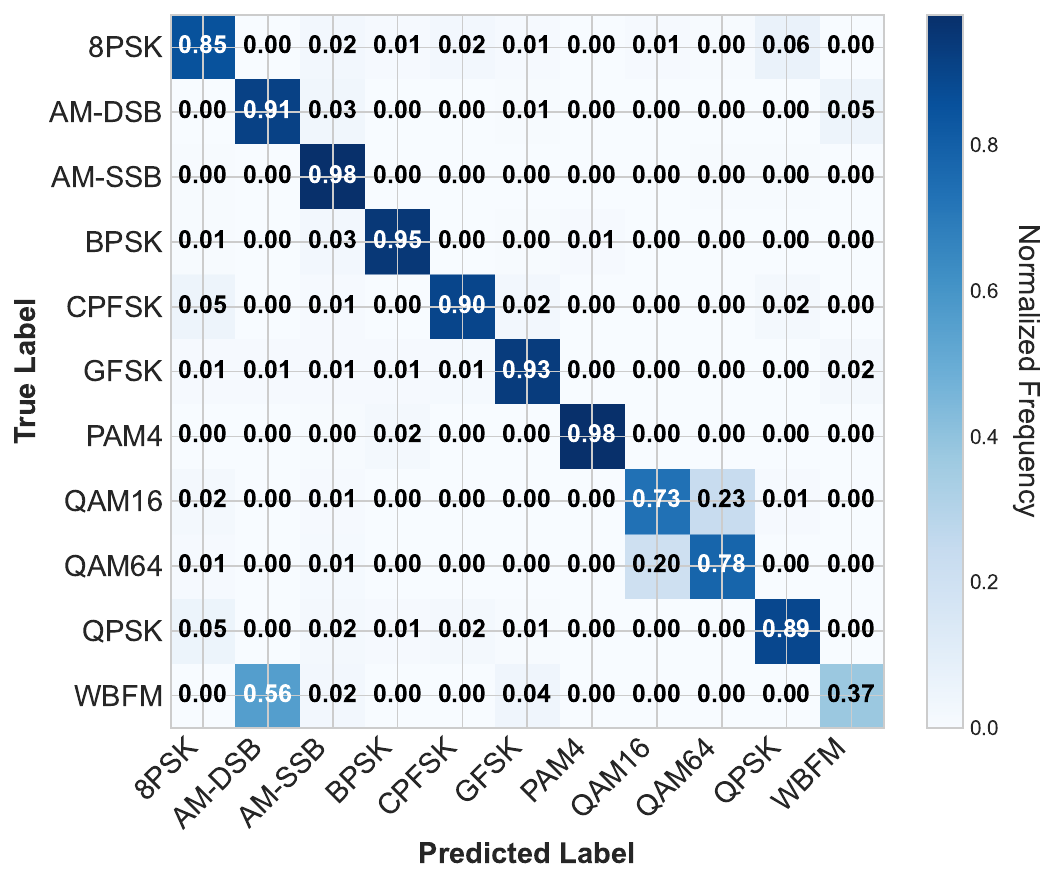}
\caption{Causal Model}
\label{fig:causal_conf}
\end{subfigure}
\hfill
\begin{subfigure}{0.3\textwidth}
\includegraphics[width=\textwidth]{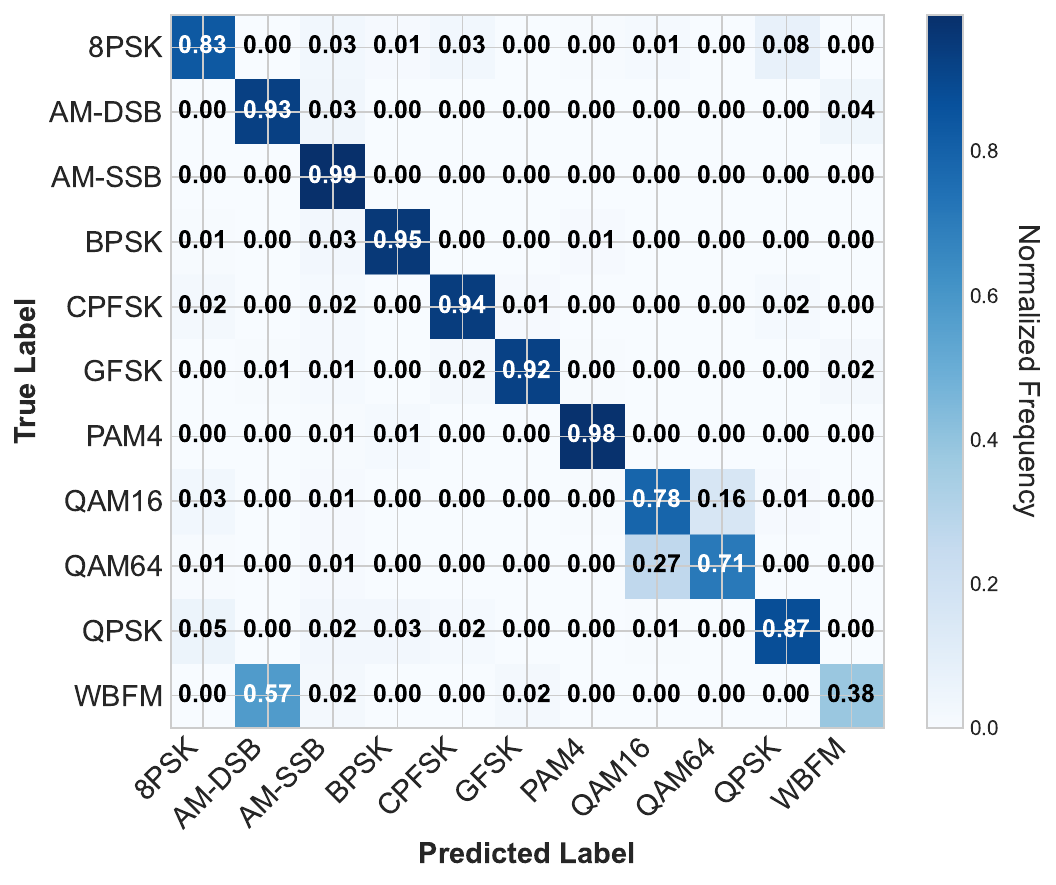}
\caption{Sparse Model}
\label{fig:sparse_conf}
\end{subfigure}
\caption{Confusion matrices for different attention mechanisms}
\label{fig:confusion_matrices}
\end{figure*}

\textbf{QAM Confusion}: QAM16 and QAM64 exhibit mutual confusion, reflecting their similar constellation structures with different symbol densities.

\textbf{AM-DSB/WBFM Confusion}: These analog modulations show cross-confusion, suggesting shared spectral characteristics that challenge all attention mechanisms.

\textbf{Digital Modulation Robustness}: BPSK, QPSK, and PSK variants achieve high classification accuracy across all attention types.

\section{Discussion and Future Work}

\subsection{Attention Mechanism Insights}

The results demonstrate that the choice of attention mechanism should be guided by the specific requirements of the application:

\textbf{Real-time Applications}: Causal and sparse attention provide excellent accuracy-efficiency trade-offs, making them ideal for real-time spectrum monitoring and cognitive radio applications.

\textbf{High-accuracy Requirements}: Baseline attention remains the best choice when maximum classification accuracy is prioritized over computational efficiency.

\textbf{Mixed Deployment}: A hybrid approach could dynamically select attention mechanisms based on detected signal characteristics.

\subsection{Architectural Considerations}

The CNN-Transformer hybrid architecture effectively combines spatial and temporal modeling capabilities. The CNN backbone provides translation-invariant feature extraction, while attention mechanisms capture long-range dependencies critical for modulation classification.

\subsection{Limitations and Future Directions}

Several limitations warrant future investigation:

\begin{itemize}
    \item \textbf{SNR Sensitivity}: Analysis across different SNR ranges could reveal attention mechanism robustness to noise.
    \item \textbf{Channel Effects}: Evaluation under realistic channel conditions (fading, multipath) would enhance practical relevance.
    \item \textbf{Adaptive Attention}: Dynamic attention pattern selection based on signal characteristics could optimize performance-efficiency trade-offs.
    \item \textbf{Larger Datasets}: Evaluation on larger, more diverse datasets would strengthen generalization claims.
\end{itemize}

\section{Conclusion}

This paper presents a comprehensive comparative analysis of attention mechanisms for automatic modulation classification. CNN-Transformer hybrid architecture with three attention variants—baseline, causal, and sparse—was evaluated on the RML2016.10a dataset. Key findings include: (1) Baseline attention achieves highest accuracy (85.05\%) but with significant computational cost, (2) Causal and sparse attention provide 83\% and 75\% inference time reductions respectively while maintaining competitive performance above 83\%, (3) Simple modulations benefit from sparse local attention while complex modulations require global context, and (4) All mechanisms struggle with analog modulations, particularly WBFM.

These results provide valuable insights for designing efficient attention-based architectures for radio signal processing applications, enabling informed trade-offs between accuracy and computational efficiency based on specific deployment requirements.

\bibliographystyle{IEEEtran}
\bibliography{ref}

\end{document}